\documentclass[prl,aps,twocolumn,floatfix,showpacs,amsmath,amssymb]{revtex4}
\usepackage{bm}
\usepackage{graphicx}

\newcommand{\be}{\begin{equation}}
\newcommand{\ee}{\end{equation}}
\newcommand{\bea}{\begin{eqnarray}}
\newcommand{\eea}{\end{eqnarray}}

\begin{document}
\def\tit#1#2#3#4#5{{#1} {\bf #2}, #3 (#4)}

\title{Spin-orbit mediated anisotropic spin interaction in interacting electron systems}
\author{Suhas Gangadharaiah, Jianmin Sun and Oleg A. Starykh}
\affiliation{Department of Physics, University of Utah, Salt Lake City, UT 84112}
\date{\today}

\begin{abstract}
We investigate interactions between spins of strongly correlated electrons
subject to the spin-orbit interaction. Our main finding is that
of a novel, spin-orbit mediated anisotropic spin-spin coupling of the van der Waals type.
Unlike the standard exchange, this interaction does not require the
wave functions to overlap. We argue that this ferromagnetic interaction is important in the
Wigner crystal state where the exchange processes are severely suppressed.
We also comment on the anisotropy of the exchange between spins
mediated by the spin-orbital coupling.
\end{abstract}
\maketitle
\label{sec:spin-spincoupling}

{\sl Introduction.} Studies of exchange interaction between localized electrons
constitutes one of the oldest topics in quantum mechanics.
Strong current interest in the possibility to control and manipulate 
spin states of quantum dots has placed this topic in the center
of spintronics and quantum computation research.
As is known from the papers of Dzyaloshinskii \cite{dzyaloshinskii58} 
and Moriya \cite{moriya60},
in the presence of the spin-orbital interaction (SOI) the exchange
is  anisotropic in spin space.

Being a manifestation of quantum tunneling, the exchange is exponentially
sensitive to the distance between electrons \cite{bernu01}. This smallness of the exchange 
leads to a large spin entropy of the
Wigner crystal state, as compared to the Fermi liquid state,
of diluted two-dimensional electron gas
in semiconductor field-effect transistors \cite{spivak04}.
The consequence of this, known as the Pomeranchuk effect,
is spectacular: Wigner crystal phase is stabilized by a finite temperature.

In this work we show that when subjected to the spin-orbit interaction,
as appropriate for the structure-asymmetric heterostructures and 
surfaces \cite{rashba84},
interacting electrons acquire a novel {\sl non-exchange} coupling between the spins.
The mechanism of this coupling is very similar to that of the
well-known van der Waals (vdW) interaction between
neutral atoms. This anisotropic interaction is of the {\sl ferromagnetic Ising} type.
It lifts extensive spin degeneracy of the Wigner crystal
and leads to the long-range ferromagnetic order.
We also re-visit and clarify the role of spin-orbit interaction
in lowering the symmetry of the exchange
coupling between spins. Particularly, we point out that
the exchange Hamiltonian, despite its anisotropic appearance,
retains spin-rotational  invariance
to the second order in the spin-orbital coupling.
We argue that spin-rotational symmetry is broken only
in the {\sl forth} order in SOI coupling.

\begin{figure}[ht]
   \includegraphics[width=1.5in]{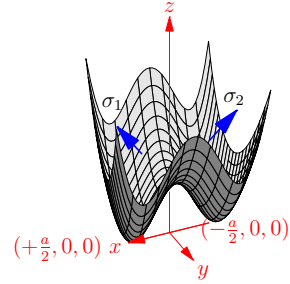}
       \caption{(color online) Two-dot potential \eqref{eq:potential}. 
       Blue (dark grey) arrows indicate electron's spins.}
  \label{fig:1}
\end{figure}

{\sl Calculation of the vdW coupling.} To illuminate the origin 
of the vdW coupling, we consider the toy problem
of two single-electron quantum dots
described by the double well potential \cite{burkard99,calderon06}, see Figure \ref{fig:1},
\begin{eqnarray}
\widetilde{V}(x_j,y_j) = \frac{m \omega^2_x}{2 a^2} (x_j^2 - \frac{a^2}{4})^2 +
\frac{m \omega^2_y}{2 }y_j^2 ,
\label{eq:potential}
\end{eqnarray}
where $\omega_{x/y}$ are confinement frequencies along $x/y$ directions.
The electrons, indexed by $j=1,2$, are subject to SOI
of the Rashba type\cite{rashba84} with coupling $\alpha_R$
\begin{eqnarray}
H_{\rm SO} = \sum_{j=1,2} \alpha_R ~\vec{p}_j \times \vec{\sigma}_j\cdot \hat{z},
\label{eq:Spin_Orbit1}
\end{eqnarray}
where $\vec{\sigma}_i$ are the Pauli matrices and $\hat{z}$ is normal to the plane
of  motion. Finally, electrons experience mutual Coulomb repulsion
so that the total Hamiltonian reads
\begin{eqnarray}
H = \sum_{j=1,2} [\frac{\vec{p}_j^2}{2m} + \widetilde{V}(x_j,y_j)]  + 
\frac{e^2}{|\vec{r}_1-\vec{r}_2|} + H_{\rm SO}.
\label{eq:Spin_Hamil00}
\end{eqnarray}
At large separation between the two dots
the exchange is exponentially suppressed and the electrons
can be treated as distinguishable particles.
One then expects that Coulomb-induced correlations in the
orbital motion of the electrons in two dots translate, via the spin-orbit interaction,
into correlation between their spins. 
Consider the distance between the dots, $a$, much greater than
the typical spread of the electron wave functions, $1/\sqrt{m \omega_x}$.
In this limit the electrons are centered about different wells,  
and the potential can be approximated as 
\begin{eqnarray}
&& V(\vec{r}_1,\vec{r}_2 ) \approx \frac{1}{2} m \omega_x^2 ((x_1- a/2)^2+ (x_2+a/2)^2) {}\nonumber\\
&&+  \frac{1}{2} m \omega_y^2 (y_1^2+ y_2^2).
\label{eq:potential1}
\end{eqnarray}
At this stage it is crucial to perform a unitary transformation \cite{raikh94,aleiner01} which
removes the linear spin-orbit term  from \eqref{eq:Spin_Hamil00}
 \be
 \label{eq:U}
 U = \exp [i m \alpha_R \hat{z}\cdot(\vec{r}_1\times
\vec{\sigma}_1  + \vec{r}_2\times \vec{\sigma}_2)].
\ee
Owing to the non-commutativity of Pauli spin matrices, SOI can not
be eliminated completely, resulting in higher order in the Rashba
coupling $\alpha_R$ contributions as given by
$\widetilde{H}=U H U^\dagger $ below
\begin{eqnarray}
&&\widetilde{H}_{\rm SO} =   \sum_{j=1,2}[-m \alpha_R^2 \tilde{L}^z_j{\tilde{\sigma}}_j^z   + \frac{4}{3}m^2 \alpha_R^3 (y_j \tilde{\sigma}^y_j +  x_j \tilde{\sigma}^x_j)\tilde{L}^z_j {} \nonumber\\
&& + \frac{2}{3} i m^2 \alpha_R^3 (y_j \tilde{\sigma}^x_j -  x_j \tilde{\sigma}^y_j)  ]
+ {\rm O}(\alpha_R^4).
\label{eq:SO_Hamilt}
\end{eqnarray}
Here $\tilde{L}^z_j$ is the angular momentum of the $j^{th}$ electron,
$\tilde{L}^z = x \tilde{p}_y - y \tilde{p}_x$, and {\sl tilde} denotes unitarily
rotated operators.
The calculation is easiest when
the confining  energy is
much greater than both the Coulomb energy $e^2/a$ and the
spin-orbit energy scale $\sqrt{m \omega}\alpha_R$.
In terms of the new (primed) coordinates
$\vec{r'}_1=\vec{r}_1-\vec{a}/2$ and $\vec{r'}_2=\vec{r}_2+\vec{a}/2$
centered about $(a/2,0)$ and $(-a/2,0)$, respectively, the interaction potential
$e^2/ |\vec{r'}_1 -\vec{r'}_2+ \vec{a}|$ is expanded  in powers of $1/a$ keeping terms
up to second order in the dimensionless relative distance $(\vec{r'}_1 -\vec{r'}_2)/a$.
The linear term, $e^2(x'_1 -x'_2)/a^2$, slightly renormalizes the
equilibrium distance between the electrons  and can be
dropped from further considerations.
In terms of  symmetric (S) and anti-symmetric (A) coordinates:
\be
x_{S/A}= \frac{x'_1 \pm x'_2}{\sqrt{2}};
\ \ \ \ \ y_{S/A}= \frac{y'_1\pm y'_2}{\sqrt{2}},
\label{eq:S/A}
\ee
the quadratic term $e^2 (2 (x'_1-x'_2)^2  - (y'_1-y'_2)^2)/2 a^3$
renormalizes the
anti-symmetric frequency
${\omega_{Ax}}^2 \rightarrow  \omega_x^2 + 4 e^2/(m a^3)$
and $\omega_{Ay}^2 = \omega_y^2 -  2 e^2/(m a^3)$
while leaving the symmetric ones unmodified,
${\omega_{Sx}}^2 =  \omega_x^2$ and $\omega_{Sy}^2 = \omega_y^2$.
Quite similarly to the textbook calculation of the vdW force \cite{griffiths},
the resulting Hamiltonian
$\widetilde{H} =\widetilde{H}_S +  \widetilde{H}_A  +   \widetilde{H}_{\rm SO}$
 becomes that of harmonic oscillators
\begin{eqnarray}
\widetilde{H}_{S/A} = \frac{\vec{\tilde{p}}^2_{S/A}}{2m} +
\frac{m}{2}(\omega^2_{xS/A} x_{S/A}^2 + \omega^2_{yS/A} y_{S/A}^2)
\label{eq:Hamiltonian_symm_anti}
\end{eqnarray}
perturbed by 
$\widetilde{H}_{\rm SO} = \widetilde{H}_{\rm SO}^{(2)} +
\delta \widetilde{H}_{\rm SO}^{(2)} + {\rm O}(\alpha_R^3)$,
where
\begin{eqnarray}
&& \widetilde{H}_{\rm SO}^{(2)} = -\frac{m \alpha_R^2}{2}
[( x_S \tilde{p}_{y S} - y_S \tilde{p}_{x S}) + S\leftrightarrow A]
(\tilde{\sigma}^z_1 +\tilde{\sigma}^z_2 ) {}\nonumber\\
&& - \frac{m \alpha_R^2}{2} [ ( x_S \tilde{p}_{y A} - y_A \tilde{p}_{x S}) +
S\leftrightarrow A  ](\tilde{\sigma}^z_1 - \tilde{\sigma}^z_2 ),
\label{eq:Spin_Orbit2b}\\
&&\delta \widetilde{H}_{\rm SO}^{(2)} = -\frac{m \alpha_R^2 a}{2\sqrt{2}}
[\tilde{p}_{y S}(\tilde{\sigma}^z_1 -\tilde{\sigma}^z_2 ) +
\tilde{p}_{y A}(\tilde{\sigma}^z_1 +\tilde{\sigma}^z_2 )]. {}
\label{eq:Spin_Orbit2a}
\end{eqnarray}

It is  evident from  Eqn. (\ref{eq:Spin_Orbit2b},\ref{eq:Spin_Orbit2a}) that the leading corrections to the ground state energy is obtained either by the excitation of a single
$y$-oscillator (through \eqref{eq:Spin_Orbit2a}) and
by the simultaneous excitation of oscillators
in both the $x$ and $y$ directions (through \eqref{eq:Spin_Orbit2b}),
\begin{equation}
 \Delta E = -\sum_{i,j=S,A}\frac{|\langle 0 | \delta\widetilde{H}_{SO}^{(2)} | 1y_{i} \rangle |^2}{ \omega_{iy}}
+\frac{|\langle 0 | \widetilde{H}_{SO}^{(2)} |1x_i 1y_{j} \rangle |^2}{\omega_{ix} + \omega_{jy}}. \nonumber
\label{eq:Pert_Hamil2}
\end{equation}
It is easy to see that the spin-dependent contributions
from $\delta\widetilde{H}_{SO}^{(2)}$
cancel exactly while those originating from
$\widetilde{H}_{\rm SO}^{(2)}$ do not, resulting in the novel spin interaction
\begin{eqnarray}
H_{\rm vdW} &=&  \frac{1}{8}m^2 \alpha_R^4
~\tilde{\sigma}_{1}^z \tilde{\sigma}_{2}^z
\Big(\phi(\omega_{Sy},\omega_{Sx}) + \phi(\omega_{Ay},\omega_{Ax}) \nonumber\\
&&- \phi(\omega_{Ay},\omega_{Sx}) - \phi(\omega_{Sy},\omega_{Ax})\Big),
\label{eq:H-vdW}
\eea
where the function $\phi$ is given by a simple expression
\be
\phi(x,y) = \frac{(x-y)^2}{ x y (x+ y)} .
\ee
In case of cylindrically symmetric dots, $\omega_x = \omega_y$,
\begin{eqnarray}
H_{\rm vdW} =  -\frac{ \alpha_R^4 e^4}{4a^6\omega_x^5}
\tilde{\sigma}_{1}^z \tilde{\sigma}_{2}^z.
\label{eq:H-vdW2}
\end{eqnarray}
The physics of this novel interaction is straightforward: it comes from the
interaction-induced correlation of the orbital motion of the two particles,
which, in turn, induces correlations between their spins via the spin-orbit
coupling.
The net Ising interaction would
have been zero if not for the shift in frequency of the
anti-symmetric mode due to the Coulomb interaction.
Note that the coupling strength exhibits 
the same power-law decay with distance
as the standard van der Waals interaction \cite{griffiths}.

From (\ref{eq:H-vdW}), it follows that in the extreme anisotropic limit of
$\omega_y \rightarrow \infty$, or equivalently, the one-dimensional
(1D) limit, there is no coupling between spins. This result is understood
by noting that 1D version of SOI, given by
$ \alpha_R \sum_j \sigma_j^y p_j^x $, can be gauged away 
to all orders in $\alpha_R$ by a unitary transformation
$U_{1D} = \exp [i m \alpha_R (x_1\sigma_1^y  + x_2\sigma_2^y)]$.
Hence the absence of the spin-spin coupling in this limit.
However, either by including magnetic field (Zeeman interaction, see below) 
in a direction
different from $\sigma^y$, or by increasing the dimensionality of the
dots by reducing the anisotropy of the confining potential, the spin-orbital
Hamiltonian acquires additional non-commuting spin operators.
The presence of the mutually non-commuting spin operators
(for example, $\sigma^x$ and $\sigma^y$ in \eqref{eq:Spin_Orbit1})
makes it impossible to gauge the SOI completely, opening
the possibility of fluctuation-generated coupling between distant spins,
as in equation \eqref{eq:H-vdW2}.

{\sl Effect of the magnetic field.}
For simplicity, we neglect orbital effects 
and concentrate on the Zeeman coupling, $H_Z = -\Delta_z \sum_j  \sigma^z_j/2$,
where $\Delta_z = g \mu_B$.
Unitary transformation \eqref{eq:U} changes it to
$H_Z - \Delta_z m \alpha_R a(\sigma^x_1 - \sigma^x_2)/2 + \delta \widetilde{H}_Z$.
Here
 \begin{eqnarray}
 \delta \widetilde{H}_Z = - \sum_{j=1,2} m \alpha_R
\Delta_z  (x'_j\tilde{\sigma}_j^x + y'_j\tilde{\sigma}_j^y)
\label{eq:Spin_Zeeman}
\end{eqnarray}
describes the coupling between the Zeeman and Rashba terms.
In the basis \eqref{eq:S/A} it reduces to
\begin{eqnarray}
\delta \widetilde{H}_Z &=&  -m\Delta_z\alpha_R \frac{y_S (\sigma^y_1 +
\sigma^y_2) + x_S
 (\sigma^x_1 + \sigma^x_2)}{\sqrt{2}}, {}\nonumber\\
&&- m\Delta_z\alpha_R \frac{y_A (\sigma^y_1 -
\sigma^y_2) + x_A
 (\sigma^x_1 - \sigma^x_2)}{\sqrt{2}}.\label{eq:Spin_Orbit4}
\end{eqnarray}
For sufficiently strong magnetic field, $\Delta_z \gg \sqrt{m\omega}  \alpha_R$,
$\widetilde{H}_{\rm SO}$ can be neglected in comparison with $\delta \widetilde{H}_Z$.
Calculating second order correction to the ground state energy of the
two dots, represented as before by $\widetilde{H}_S + \widetilde{H}_A$,
and extracting the spin-dependent contribution, we obtain
\be
\Delta E_Z =-\Delta_z^2 \alpha_R^2\frac{e^2}{a^3}(2 \frac{\sigma^x_1
\sigma^x_2 }{\omega_{x}^4} - \frac{\sigma^y_1
\sigma^y_2}{\omega_{y}^4}). 
\label{eq:delta-Z}
\ee
In the extreme
anisotropic limit $\omega_y \rightarrow \infty$ the dots become
1D and we recover the result of Ref. \onlinecite{flindt06}.
For the isotropic limit $\omega_x =\omega_y$, the
coupling of spins acquires a magnetic dipolar structure identical to
that found in Ref. \onlinecite{trif07}.

{\sl Anisotropy of the exchange.}
Next, we allow for the electron tunneling between the dots.
The spin dynamics of the electrons is now described by the sum
of exchange and the van der Waals interactions,
$H=H_{\rm Ex} +  H_{\rm vdW}.$
Here  the exchange coupling, $H_{\rm Ex}$, contains both isotropic and possible anisotropic
interactions, while $H_{\rm vdW}$ is given by \eqref{eq:H-vdW} and \eqref{eq:H-vdW2}.
In the absence of spin-orbit interaction, the total spin is conserved and the Hamiltonian is SU(2)
invariant.  As such, the only spin interaction
allowed has the well known isotropic  form
$H_{\rm Ex} \sim \vec{\sigma}_1 \cdot \vec{\sigma}_2$.
The anisotropy of the exchange
is mediated by the spin-rotational symmetry breaking SOI \eqref{eq:Spin_Orbit1}.
When the tunneling is no longer spin-conserving,
electron spins precess while
exchanging their respective positions, giving rise to the anisotropic terms.
As a result \cite{kavokin04,stepanenko03,imamura04}
\be
H_{\rm Ex}= \frac{J}{4}\Big( b \vec{\sigma}_1\cdot\vec{\sigma}_2 +
D\hat{d} \cdot \vec{\sigma}_1\times \vec{\sigma}_2 +
\Gamma (\hat{d}\cdot\vec{\sigma}_1)(\hat{d}\cdot\vec{\sigma}_2)\Big),
\label{eq:exchange1}
\ee
where  $\hat{d}$ is the unit Dzyloshinskii-Moriya vector, of amplitude $D$, with
odd dependence on the spin-orbit coupling $\alpha_R$.
Coefficients $b$ and $\Gamma$ have  even dependence
on the spin-orbit coupling \cite{moriya60,aharony92}, while the exchange integral
$J$, independent of  $\alpha_R$ in this representation, sets the overall energy scale.
The direction of the DM vector can be understood as follows.
As the $D$-term must be even under exchange operation $\mathbb{P}: 1\leftrightarrow 2$,
its amplitude must be odd with respect to inter-spin distance $\vec{a}=\vec{r}_1-\vec{r}_2=a \hat{x}$,
hence $\hat{d} \sim \hat{a}=\hat{x}$. In addition, as $\hat{z}\to -\hat{z}$ transformation
in \eqref{eq:Spin_Orbit1} changes sign of $\alpha_R$, it must be that
$\hat{d} \sim \hat{z}$ as well. Thus, it must be that $\hat{d} = \hat{z} \times \hat{a} = \hat{y}$.

In the simplest approximation one neglects 
the ``remnants'' of SOI \eqref{eq:SO_Hamilt}
altogether and writes the only possible exchange coupling
$\widetilde{H}_{\rm Ex }^{(0)}= \frac{J}{4}\vec{ \tilde{\sigma}}_1 \cdot \vec{ \tilde{\sigma}}_2$
in terms of {\sl unitarily transformed} 
spin operators $\vec{\tilde{\sigma}}_j$.
The meaning of this interaction is understood in the original basis
by undoing the unitary transformation, $H_{\rm Ex }^{(0)} = U^\dagger \widetilde{H}_{\rm Ex}^{(0)}U$.
Using \eqref{eq:U} and replacing $\vec{r}_1,~\vec{r}_2$ by their
respective average values, $a/2 \hat{x}$ and $-a/2 \hat{x}$, 
one observes that 
spin $1$ ($2$) is rotated about $\hat{y}$ axis by the angle $\theta = m\alpha_R a$
in clockwise (counterclockwise) direction. As a result, one immediately
obtains Eq.\eqref{eq:exchange1} with parameters
\be
b_0 = \cos2\theta , D_0 = \sin2\theta , 
\Gamma_0 = 1 -  \cos2\theta ,  \hat{d} = \hat{y}.
\label{eq:exchange2}
\ee
As it originated from the $SU(2)$-invariant 
scalar product $\vec{ \tilde{\sigma}}_1 \cdot \vec{ \tilde{\sigma}}_2$,
the Hamiltonian \eqref{eq:exchange1} with parameters \eqref{eq:exchange2} does {\sl not}
break spin-rotational SU(2) symmetry, despite its asymmetric appearance.
Because of its ``non-diagonal" nature, the $D$-term affects the eigenvalues
only in $D^2 \sim \theta^2$ order, and must always be considered {\sl together} with
the $\Gamma$-term. In the current situation \eqref{eq:exchange2}, the
two contributions compensate each other {\sl exactly}.
This important observation, made in Ref.\onlinecite{aharony92}
(see also \cite{zheludev99}), was overlooked in several recent 
calculations of the DM term \cite{kavokin01,gorkov03,kavokin04}.

It is thus clear that the symmetry-breaking DM term must
originate from so far omitted $\widetilde{H}_{\rm SO}$ \eqref{eq:SO_Hamilt}.
To capture it, we set up the exchange problem calculation along the lines
of the standard Heitler-London (HL) approach. Despite its well-known
shortcomings \cite{herring62,herring64,gorkov64}, this approach offers
conceptually simple way to estimate exchange splitting \cite{calderon06} 
and the structure of anisotropic spin coupling.
Our basis set is formed by the antisymmetrized two-particle wave function
$|\widetilde{\psi}\rangle = |\psi\rangle - \mathbb{P} |\psi\rangle$,
\begin{eqnarray}
|\psi\rangle = \varphi(1,2)\{c_1|\uparrow \uparrow\rangle +
c_2|\uparrow \downarrow\rangle +
c_3|\downarrow \uparrow\rangle  + c_4|\downarrow \downarrow\rangle\}
\label{eq:antisymmetric-state}
\end{eqnarray}
is written in terms of unknown coefficients $c_{1-4}$.
Here $\varphi(1,2) = f(x_1-a/2) f(y_1)  f(x_2+a/2) f(y_2)$
describes spatial wave function of distinguishable particles localized near $(a/2,0)$ and
$(-a/2,0)$, respectively, and $f(x-x_0)$ denotes the ground state wave function of
one-dimensional harmonic oscillator centered around $x=x_0$. As constructed,
$\varphi(1,2)$ is the lowest energy eigenstate of two particles moving in the potential
profile \eqref{eq:potential1}.

The rest of the confining potential, Eq.\eqref{eq:potential}, {\sl together with}
the SOI \eqref{eq:SO_Hamilt}, forms the perturbation
\be
V_{\rm pert}(1,2) = \sum_{j=1,2}\widetilde{V}(x_j,y_j) - V(\vec{r}_1,\vec{r}_2) +
\widetilde{H}_{\rm SO} ,
\ee
which is responsible for removing spin degeneracy 
of states contributing to \eqref{eq:antisymmetric-state}.
The eigenvalue problem
\be
(H_0 + V_{\rm pert}) |\widetilde{\psi}\rangle = E |\widetilde{\psi}\rangle ,
\label{eq:eigenvalue1}
\ee
where $H_0$ is the sum of kinetic energy and confinement potential \eqref{eq:potential1},
is formulated as a $4\times4$ matrix problem by multiplying \eqref{eq:eigenvalue1}
by the bra $\langle s_1 s_2| \varphi(1,2)$ from the left (here $s_{j=1,2} = \uparrow$ or $\downarrow$)
and integrating the result over the whole  space.
The  obtained exchange Hamiltonian
for the {\sl rotated} spins $\vec{\tilde{\sigma}}$ is
of the form \eqref{eq:exchange1} with
\be
J = \frac{3}{2} m\omega_x^2 a^2 e^{-m\omega_x a^2/2} ,
D = \frac{32 m \alpha_R^3}{9 \omega_x \omega_y a} , 
\label{eq:HLparameters}
\ee
while $b=1, \Gamma=0$ to this order.
The calculation sketched is valid in the large separation limit, $a\gg1/\sqrt{m\omega_x}$,
and its most important feature is the scaling $D \sim \alpha_R^3$ between the
DM coupling and the spin-orbital one. This result is due to the fact that
$O(\alpha_R^2)$ term in \eqref{eq:SO_Hamilt} excites both $x$ and $y$ oscillators.
 Since the wave function \eqref{eq:antisymmetric-state}
contains only the ground states of the oscillators, the $O(\alpha_R^2)$ term
drops out and the first asymmetric correction originates in $O(\alpha_R^3)$ terms
of \eqref{eq:SO_Hamilt}. We checked that this crucial feature
is not an artifact of the HL approximation and is  also obtained from a more reliable
``median-plane" approach \cite{herring64,gorkov64,efros99,gorkov03},
which we initiated. 

Noting that the DM term $D\hat{y}\cdot\vec{\tilde{\sigma}}_1 \times \vec{\tilde{\sigma}}_2$
affects the eigenvalue of the two-spin problem only in $D^2$ order, we conclude that exchange
asymmetry due to the spin-orbit interaction may appear only in $\alpha_R^4$ or higher order.
This is because the effect of $\Gamma$-term in  \eqref{eq:exchange1} on the eigenvalues
is of first order in $\Gamma$, and our calculation shows that $\Gamma \sim O(\alpha_R^4)$.
Being proportional to $J$, see \eqref{eq:HLparameters}, 
this contribution is also exponentially small.
We then conclude that  the leading source of spin anisotropy
is provided by the vdW contribution \eqref{eq:H-vdW} and \eqref{eq:H-vdW2},
which does not contain an exponential smallness of the exchange.

{\sl Estimate of the vdW coupling.}
We now turn our attention to physical manifestations of the vdW spin coupling
in the Wigner crystal. 
Neglecting the exchange interaction for the moment, we consider a two-electron
problem within the frozen lattice approximation in which all other electrons are
assumed fixed in their equilibrium lattice positions.  The potential energy then
is just that of four harmonic oscillators \cite{flambaum99}
with frequencies $\omega_{\xi,\eta} = \sqrt{(\gamma \mp 2)/(m^2 a_B a^3)}$
and $\omega_{u,v} = \sqrt{(\gamma \mp 1)/(m^2 a_B a^3)}$, 
in notations of Ref.\onlinecite{flambaum99}.
Here  $\gamma\approx 5.52$,\cite{flambaum99}
$a_B = \kappa/(m e^2)$ is the Bohr radius, $\kappa$ is the dielectric constant
and $a$ is the lattice constant of the electron crystal, inversely
proportional to the electron density $n$: $a=(2/\sqrt{3}n)^{1/2}$.
Repeating the steps that led to \eqref{eq:H-vdW} we obtain for the Wigner crystal problem
\be
H_{\rm vdW}^{\rm wigner} =  m^2\alpha_R^4 B ~\tilde{\sigma}_{1}^z \tilde{\sigma}_{2}^z
= g_{\rm vdW} ~\tilde{\sigma}_{1}^z \tilde{\sigma}_{2}^z
\label{eq:wigner}
\ee
where $B = [\phi(\omega_\xi,\omega_v) + \phi(\omega_\eta,\omega_u) -
\phi(\omega_\xi,\omega_u) - \phi(\omega_\eta,\omega_v)]/8 = -3.75\cdot 10^{-3} \sqrt{m^2 a_B a^3}$.
The spin-orbit mediated ferromagnetic coupling removes extensive spin degeneracy
of the crystal, suppressing the Pomeranchuk effect physics \cite{spivak04}.
Being of non-frustrated nature, it establishes long-range magnetic
order of Ising type with the ordering temperature of the order of the vdW constant
$g_{\rm vdW}$ \eqref{eq:wigner}. 
It should be compared with the much studied Heisenberg exchange
$J_{\rm wc} = c(r_s) \exp[- 1.612 ~\sqrt{r_s}]$, expressed 
in Rydbergs ${\cal{R}}=1/(2 m a_B^2)$.
Here $r_s = 1/\sqrt{\pi a_B^2 n}$ is the dimensionless measure of the interaction strength,
and the pre-factor $c(r_s)$ is a smooth function of it \cite{bernu01}.
We find that $g_{\rm vdW}$ dominates the exchange for $r_s > r_s^* \approx 20$
in  InAs, which has $\alpha_R \approx 1.6 \cdot 10^{4}$m/s \cite{grundler}.
For GaAs, with $\alpha_R \approx 300$m/s  \cite{weber}, more diluted
situation is required, $r_s^* \approx 90$.
Given that multi-particle ring-exchange processes on
the triangular lattice strongly frustrate any ordering tendencies
due to the exchange \cite{bernu01}, it appears that our estimate is just a lower bound
on the critical density below which spin-orbit-induced ferromagnetic state should be expected.

We would like to thank L. Balents, L. Glazman, D. Maslov,
E. Mishchenko, M. Raikh, O. Tchernyshyov, and, 
especially, K. Matveev, for productive discussions.
O.A.S. is supported by ACS PRF 43219-AC10.

\end{document}